\documentclass{aastex62}

\begin{document}

\title{Can We Distinguish the Source Region Location of Filament/Prominence Eruptions from the Sun-as-a-star H$\alpha$ Spectrum?}

\correspondingauthor{Yijun Hou}
\email{yijunhou@nao.cas.cn}

\author[0009-0004-4079-1342]{Junyi Zhang}
\affiliation{State Key Laboratory of Solar Activity and Space Weather, National Astronomical Observatories, Chinese Academy of Science, Beijing 100101, People’s Republic of China}
\affiliation{School of Astronomy and Space Science, University of Chinese Academy of Sciences, Beijing 100049, People’s Republic of China}

\author[0000-0002-9534-1638]{Yijun Hou}
\affiliation{State Key Laboratory of Solar Activity and Space Weather, National Astronomical Observatories, Chinese Academy of Science, Beijing 100101, People’s Republic of China}
\affiliation{School of Astronomy and Space Science, University of Chinese Academy of Sciences, Beijing 100049, People’s Republic of China}

\author[0000-0002-3657-3172]{Xiaofeng Liu}
\affiliation{State Key Laboratory of Solar Activity and Space Weather, National Astronomical Observatories, Chinese Academy of Science, Beijing 100101, People’s Republic of China}
\affiliation{Key Laboratory of Dark Matter and Space Astronomy, Purple Mountain Observatory, Chinese Academy of Sciences, Nanjing 210023, People’s Republic of China}
\affiliation{School of Astronomy and Space Science, University of Science and Technology of China, Hefei 230026, People’s Republic of China}

\author[0000-0001-6655-1743]{Ting Li}
\affiliation{State Key Laboratory of Solar Activity and Space Weather, National Space Science Center, Chinese Academy of Sciences, Beijing 100190, China}
\affiliation{State Key Laboratory of Solar Activity and Space Weather, National Astronomical Observatories, Chinese Academy of Science, Beijing 100101, People’s Republic of China}
\affiliation{School of Astronomy and Space Science, University of Chinese Academy of Sciences, Beijing 100049, People’s Republic of China}

\author[0000-0002-7544-6926]{Shihao Rao}
\affiliation{School of Astronomy and Space Science, Nanjing University, Nanjing 210023, People’s Republic of China}
\affiliation{Key Laboratory of Modern Astronomy and Astrophysics (Nanjing University), Ministry of Education, Nanjing 210023, People’s Republic of China}

\author[0000-0002-1190-0173]{Ye Qiu}
\affiliation{Institute of Science and Technology for Deep Space Exploration, Suzhou Campus, Nanjing University, Suzhou 215163, People’s Republic of China}

\author{HuiPing Jin}
\affiliation{China West Normal University, Sichuan 637009, People’s Republic of China}
\affiliation{State Key Laboratory of Solar Activity and Space Weather, National Astronomical Observatories, Chinese Academy of Science, Beijing 100101, People’s Republic of China}

\author[0009-0007-4469-0663]{Yingjie Cai}
\affiliation{State Key Laboratory of Solar Activity and Space Weather, National Astronomical Observatories, Chinese Academy of Science, Beijing 100101, People’s Republic of China}
\affiliation{School of Astronomy and Space Science, University of Chinese Academy of Sciences, Beijing 100049, People’s Republic of China}

\author{Yangrui Chen}
\affiliation{Southwest Jiaotong University, Sichuan 611756, People’s Republic of China}
\affiliation{State Key Laboratory of Solar Activity and Space Weather, National Astronomical Observatories, Chinese Academy of Science, Beijing 100101, People’s Republic of China}

\author[0000-0001-7693-4908]{Chuan Li}
\affiliation{School of Astronomy and Space Science, Nanjing University, Nanjing 210023, People’s Republic of China}
\affiliation{Key Laboratory of Modern Astronomy and Astrophysics (Nanjing University), Ministry of Education, Nanjing 210023, People’s Republic of China}
\affiliation{Institute of Science and Technology for Deep Space Exploration, Suzhou Campus, Nanjing University, Suzhou 215163, People’s Republic of China}

\begin{abstract}

Solar filament/prominence eruptions can significantly perturb geospace when originating from favorable source locations and directions. While stellar analogs have been recently reported, the disk locations and magnetic environments of their source regions remain spatially unresolved on other stars. To bridge this gap, we investigate the typical Sun-as-a-star H$\alpha$ temporal spectral characteristics of solar filament/prominence eruptions with different source region locations (on-disk vs. limb, active region vs. quiet-Sun region). It is revealed that limb eruptions are characterized by blueshifted/redshifted emission caused by the bright off-limb erupting structures, whereas on-disk eruptions may show blueshifted absorptions due to the dark erupting filaments. Among the limb eruptions, front-side limb eruptions usually display line center emission before the blueshifted/redshifted emission, while far-side limb eruptions show the opposite sequence. Moreover, the magnetic environment at source also shapes the spectral characteristics. On-disk filament eruptions from active region exhibit much more intense flare-ribbon-dominated line center emission features compared with those from quiet-Sun region. Limb active region eruptions often show single-wing emissions, whereas large-scale quiet-Sun region (quiescent) prominence eruptions frequently display expansion-induced emission in both wings followed by line center absorption due to the disappearance of bright prominence. These distinct Sun-as-a-star H$\alpha$ spectral characteristics, dependent on eruption location, provide a diagnostic basis for inferring source regions of stellar filament/prominence eruptions from spatially unresolved H$\alpha$ spectra.

\end{abstract}


\section{Introduction}\label{1}

Solar filaments (or prominences when observed at the solar limb) are cool, dense plasma structures suspended in the hot corona by magnetic fields \citep[e.g.,][]{2010SSRv..151..333M,2020RAA....20..166C}. They are fundamental structures of the solar atmosphere and are frequently associated with the storage of free magnetic energy \citep{2014masu.book.....P,2014LRSP...11....1P}. The eruption of a filament/prominence, often accompanied by solar flares and coronal mass ejections (CMEs) \citep{2011LRSP....8....1C}, can release enormous amounts of energy, plasma, and embedded magnetic field into interplanetary space. These eruptions are primary drivers of geomagnetic storms and can affect satellite operations, communications, and power grids \citep{2022Atmos..13.1781G}. Previous studies have demonstrated that the heliographic locations of the intense-storm-causing solar eruptions are concentrated near the solar disk
center, with a slight bias to the western hemisphere \citep[e.g.,][]{1943MNRAS.103..244N,2011JGRA..116.4104W,2015AdSpR..55.2745S}. The eruption's initial propagation direction is also critical to assessing its potential geoeffectiveness, which is closely related to the magnetic environment of eruption source region \citep[e.g.,][]{2020JGRA..12527530W,2023ApJ...953...68L,2025ApJ...978...41S}. Therefore, understanding the initiation, specifically the on-disk location and magnetic environment of source region, of filament/prominence eruptions is a core objective in both solar physics and space weather forecasting.

Solar filaments and prominences are broadly classified into two types based on their location and magnetic environment: active region filaments/prominences and quiet-Sun region filaments/prominences. These two types exhibit distinct physical properties and eruptive behaviors. Active region filaments/prominences are smaller in spatial scale, lower in height, and embedded within complex, strong magnetic fields of active regions \citep[e.g.,][]{2015ASSL..415.....V}. Their eruptions are often directly triggered by magnetic reconnection in the low corona, involved in intense flares and fast, impulsive CMEs \citep[e.g.,][]{2018A&A...619A.100H,Hou2023Apj,2025ApJ...984....4W}. In contrast, quiescent filaments/prominences are larger, higher, and reside in the quiet Sun with simpler, weaker magnetic arcades \citep[e.g.,][]{1995ApJ...443..818L,2020A&A...640A.101H,2025ApJ...981..139Y}. Their eruptions are typically more gradual and are often associated with slower, massive CMEs \citep[e.g.,][]{2014LRSP...11....1P,2021ApJ...914...39L}. It is obvious that the observed characteristics and associated phenomena of a filament/prominence eruption are highly dependent on its source location and the surrounding magnetic field configuration.

Beyond the Sun, filaments/prominences are believed to exist on other magnetically active stars \citep[e.g.,][]{2022NatAs...6..241N,2024ApJ...961...23N,2026NatAs..10...64N,2025ApJ...978L..32L,2025ApJ...979...93K}. However, due to observational limitations, stellar filaments/prominence studies are far less extensive than solar studies. While stellar flares are routinely observed in photometric light curves \citep[e.g.,][]{2024LRSP...21....1K}, the associated coronal mass ejections (stellar CMEs) and filament/prominence eruptions are difficult to capture directly. Unlike the Sun, where we can pinpoint the exact source location of the solar eruption, direct imaging with spatial resolution is difficult for other stars. This severely hinders our understanding of the source regions and the underlying eruption mechanisms on these distant unresolved stars.

To bridge the gap between spatially resolved solar observations and unresolved stellar data, “Sun-as-a-star” analysis has emerged as a crucial methodology \citep{2016SoPh..291.1761H}. By integrating solar full-disk observations to mimic a point star source, we can test whether specific spectral signatures can be used to diagnose the physical properties of stellar eruptions \citep[e.g.,][]{2022ApJ...939...98O,2022ApJ...933..209N}. Recently, through separately analyzing the subregions dominated by different dynamical processes, \citet{2025ApJ...993..126L} obtained typical Sun-as-a-star spectral characteristics of different physical process during solar eruptions. Significant progress has been made in detecting and analyzing solar eruptions in Sun-as-a-star spectra \citep[e.g.,][]{2022ApJS..260...36Y,2022ApJ...931...76X,2024ApJ...964...75O,2024ApJ...966...45M,2024ApJ...974L..13O,2024A&A...682A..46P,2026ApJ...997..242C}. However, a critical question remains yet largely unexplored: whether the source location of a filament/prominence eruption event (e.g., limb vs. on-disk, or active regions vs. quiet-Sun regions) can be inferred solely from the integrated H$\alpha$ spectra. Solving this would significantly enhance our ability to interpret stellar H$\alpha$ spectra of similar stellar eruptions.

In this paper, we utilize high-resolution spectroscopic observations from the Chinese H$\alpha$ Solar Explorer (CHASE) \citep{2022SCPMA..6589602L} to analyze a series of filament and prominence eruption events. By employing Sun-as-a-star analysis techniques on the spatially resolved H$\alpha$ spectral data, we aim to explore whether integrated spectral profiles contain information that can diagnose the source location of the erupting filament or prominence. The ultimate goal is to develop a method that can be applied to stellar observations to constrain the location of stellar filament/prominence eruptions and, by extension, their source magnetic environments. The paper is organized as follows: Section \ref{2} describes the CHASE observations and the data analysis methods. Section \ref{3} presents the analysis of these events and discusses the results. Section \ref{4} summarizes our conclusions.

\section{Observations and Methods}\label{2}
\subsection{Overview of Events}

In this study, we selected 7 solar filament/prominence eruption events observed by the H$\alpha$ Imaging Spectrograph (HIS) on board CHASE (see in Table \ref{table1}). These events were chosen to represent a diverse range of eruption source locations on the solar disk or off the limb. The classification of these events is as follows: Events 1 and 4 are on-disk active region filament eruptions; Event 2 is a front-side limb active region filament eruption; Event 3 is a far-side limb active region filament eruption; Event 5 is an on-disk quiet-Sun region filament eruption; and Events 6 and 7 are limb quiet-Sun region filament (quiescent prominence) eruptions. Some information of these events is presented in Table \ref{table1} and details are shown in Section \ref{3}.

\begin{deluxetable*}{lcccc}
\tablecaption{Information of Selected Filament/Prominence Eruption Events \label{table1}}
\tablecolumns{5}
\tablewidth{0pt}
\tablehead{
\colhead{Event} & \colhead{Date (UT)} & \colhead{GOES Class} & \colhead{Source Location} & \colhead{Magnetic Environment}
}
\startdata
1 & 2024 Jul 29 &M8.7 & On-disk Eruption & Active Region Eruption\\
2 & 2024 Apr 13 &M2.4 & Front-side Limb Eruption & Active Region Eruption\\
3 & 2024 Jun 10 &X1.6 & Far-side Limb Eruption & Active Region Eruption\\
4 & 2024 May 5 &M7.5 & On-disk Eruption & Active Region Eruption\\
5 & 2024 Apr 11 &/ & On-disk Eruption & Quiet-Sun Region Eruption\\
6 & 2025 Aug 20 &/ & Limb Eruption & Quiet-Sun Region Eruption\\
7 & 2025 Sep 30 &/ & Limb Eruption & Quiet-Sun Region Eruption\\
\enddata
\end{deluxetable*}

The CHASE/HIS instrument provides full-disk spectroscopic observations of the Sun in the H$\alpha$ (6559.7-6565.9 \AA) and the Fe I (6567.8-6570.6 \AA) waveband. The data used in this work underwent standard preprocessing steps, including flat-field, dark-field, and slit-image-curvature corrections, as described in previous instrument calibration reports \citep{2022SCPMA..6589603Q}. The instrument routinely provides full-disk scanning spectra with a high spectral resolution of 0.048 \AA\ per pixel. To provide context for the energetic output of these eruptions, particularly the associated solar flares, we also utilized soft X-ray flux data (1-8 \AA) from the Geostationary Operational Environmental Satellite (GOES).

\subsection{Sun-as-a-star Analysis}

In this paper, we use the “virtual Sun-as-a-star method”, which assumes that the variations of Sun-as-a-star line profiles are only caused by specific regions of interest (target regions, TRs), to investigate whether the integral spectral signature can reveal information of eruption locations. By summing the spectral profiles over TRs and normalizing the intensity, we obtain a single time-series spectrum:
\begin{equation}
\Delta S(t, \lambda ) = \frac{\frac{\int_{TR} I(t, \lambda ,x ,y)dxdy}{\int_{TR} I(t, \lambda _{cont} ,x ,y)dxdy}\times \int_{TR} I(t_{0}, \lambda _{cont},x ,y) dxdy-\int_{TR} I(t_{0}, \lambda ,x ,y)dxdy}{\int_{full\  disk} I(t_{0}, \lambda _{cont},x ,y) dxdy},
\end{equation}
where $I(t, \lambda ,x ,y)$ is the intensity function of the observed time $t$, the wavelength $\lambda $, and the spatial position $(x,y)$, $t_0$ is a pre-event time and $\lambda_{cont}$ is a continuum wavelength in the spectral window of Fe I.

This approach allows us to directly compare the integrated spectral evolution with the spatially resolved dynamics of the filaments and prominences. We also obtain the differential equivalent width ($\Delta$EW): $\Delta EW = \int_{H\alpha -\Delta \lambda}^{H\alpha +\Delta \lambda}\Delta S(t,\lambda)d\lambda$, where $\Delta \lambda$ represents the integration range around H$\alpha$ line center. The details and discussion of the mathematical formulation follow the methodology established in previous works \citep[e.g.,][]{2022ApJ...939...98O, 2025ApJ...993..126L}. In this study, we adopt the formalism of \citet{2025ApJ...993..126L} because our space-based CHASE data do not require quiet-region normalization to correct for terrestrial atmospheric effects.

\subsection{Doppler Velocity Field Calculation}

To support and validate our Sun-as-a-star analysis and quantitatively analyze dynamical processes of erupting filaments/prominences in different source regions, we calculated the Doppler velocity field of the TRs using single- and two-cloud model \citep{2024ApJ...961L..30Q}. From the single- or two-cloud model, we can get the Doppler velocity in the radiative transfer equation via the optical depth. We can calculate the parameters in the cloud model by fitting the H$\alpha$ line profiles. For each pixel, we first apply the single-cloud model. If the fitting result is not satisfactory, we then adopt the two-cloud model, which yields two velocities. We compare these two velocities with the single‑cloud result, and the velocity that is closer to the single‑cloud value is taken as the velocity of the main plasma body. In our Doppler velocity maps, most pixels come from the single-cloud model, while a small fraction come from the main‑body cloud of the two-cloud model. The details and discussion of the cloud model, including the specific centroiding algorithm and the calibration of the reference wavelength, was described in \citet{2024ApJ...961L..30Q}. This processing results in velocity maps that are co-temporal with the intensity images, enabling a detailed study of the kinematic evolution of the filaments and prominences during the eruption phase.

\section{Results and Discussion}\label{3}

The primary objective of this study is to investigate whether Sun-as-a-star spectral signatures can be effectively used to diagnose the source location of solar filament and prominence eruptions. Specifically, we aim to discern how the integrated H$\alpha$ profiles differ depending on whether the eruption originates from the solar disk or the limb, and whether it arises from an active region or a quiet-Sun region. To achieve this, we performed a systematic comparative analysis of the 7 selected solar filament/prominence eruption events. By grouping these events based on their source region locations, we explore distinct spectral characteristics that may serve as diagnostic tools for unresolved stellar observations. The following subsections detail these comparisons, focusing on distinguishing between on-disk and limb events, front-side and far-side limb events, and eruptions originating from different magnetic environments (active regions vs. quiet-Sun regions).

\subsection{On-disk vs. Limb Active Region Eruptions}

\begin{figure}[h]
\centering
\includegraphics [width=0.99\textwidth]{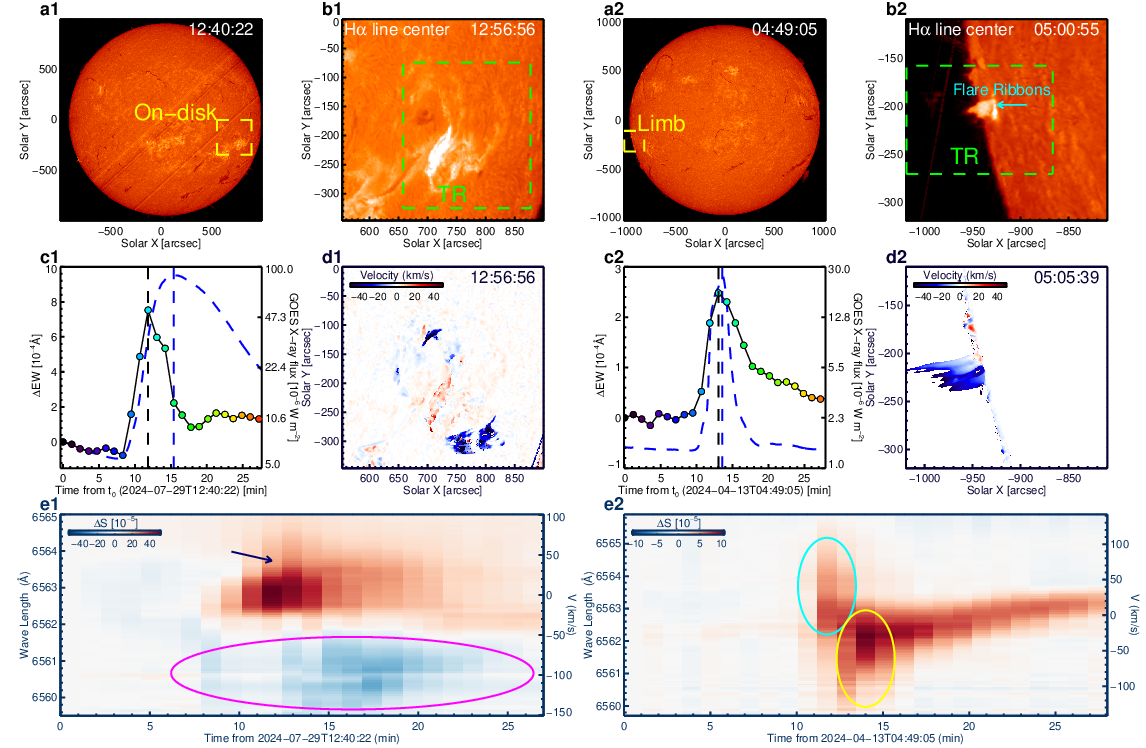}
\caption{CHASE imaging and spectral observations of the events occurred on 2024 Jul 29 (event 1, on-disk active region eruption) and 2024 Apr 13 (event 2, front-side limb active region eruption). (a1) The full-disk solar imaging of event 1 at the  H$\alpha$ line center. The yellow dashed region is the cropped source region as (b1). (b1) The source region of event 1. The green dashed region is the target region (TR). (c1) Light curves of H$\alpha$ $\Delta$EW (colored circles) and GOES soft X-ray 1–8 \AA\ flux (blue dashed curve) of event 1. The black and blue vertical dashed lines mark the peak times of $\Delta$EW and soft X-ray. (d1) The Doppler velocity filed of the source region of event 1. (e1) The time series of Sun-as-a-star H$\alpha$ dynamic spectrum in the source region of event 1. The navy arrow marks the red asymmetry. (a2)-(e2) The same as (a1)-(e1) but for event 2. The navy and cyan arrow marks the flare ribbons. An animation is available online, which displays the evolution of events 1 and 2. The animation’s duration is 2.5 s.}
\label{fig1}
\end{figure}

Figure \ref{fig1} and the associated online animation present an on-disk active region filament eruption (event 1) and a front-side limb active region filament eruption (event 2). Event 1 occurred on July 29, 2024, accompanied by an M8.7 flare. According to GOES records, the flare started at 12:47 UT, ended at 13:04 UT, and peaked at 12:55 UT. The spatially integrated spectral characteristics of the source region mainly exhibit signatures typical of flare ribbons (see Figure \ref{fig1}(e1)), namely enhanced emission at the H$\alpha$ line center and an increase in line width, with an overall roughly symmetric profile. A red asymmetry (see the navy arrow in Figure \ref{fig1}(e1)) is detected around the peak time, which is commonly interpreted as evidence of chromospheric condensation \citep{2022NatAs...6..241N}. During the subsequent decay phase, the emission intensity gradually decreases while the line profile remains symmetric. An absorption signal is observed in the blue wing (see the magenta ellipse in Figure \ref{fig1}(e1)), which is caused by the obscuration of the erupting filament. This indicates that the filament possessed an initial velocity component directed toward Earth at the onset of its eruption. Stripe-like noise signals in the blue wing caused by Si I and Ti II lines are weak and have negligible impact on our results (see more details in Figure 5 of \citet{2025ApJ...993..126L}). The $\Delta$EW peaks before the GOES soft X-ray peak (see Figure \ref{fig1}(c1)), showing a significant enhancement.

Event 2 occurred on April 13, 2024, accompanied by an M2.4 flare. According to GOES, the flare started at 04:58 UT, ended at 05:06 UT, and peaked at 05:02 UT. The spatially integrated spectral characteristics of the source region reveals emission signals at the line center, indicative of flare ribbons, as well as a red asymmetry (see the cyan arrow in Figure \ref{fig1}(e2)). However, this red asymmetry is short-lived. Subsequently, an emission signal emerges in the blue wing (see the yellow ellipse in Figure \ref{fig1}(e2)) and gradually drifts toward the red wing. This signal corresponds to off-limb erupting prominence material, suggesting that this prominence possessed a velocity component directed toward the observer during its eruption, before subsequently falling back onto the Sun. The $\Delta$EW shows a significant enhancement, peaking before the GOES SXR peak (see Figure \ref{fig1}(c2)). This temporal relationship differs from the findings mentioned in \citet{2022ApJ...939...98O}, where the prominence eruption signal typically peaks after the SXR peak. This discrepancy likely arises because our integration region for the Sun-as-a-star analysis includes contributions from the visible flare ribbons (see the cyan arrow in Figure \ref{fig1}(b2)). Additionally, during the late stage of the eruption, small-scale post-flare loops emerged in the northern part of the TR. These loops, however, are too small and too weak to have a noticeable impact on our results dominated by the erupting filament.

\begin{figure}[h]
\centering
\includegraphics [width=0.9\textwidth]{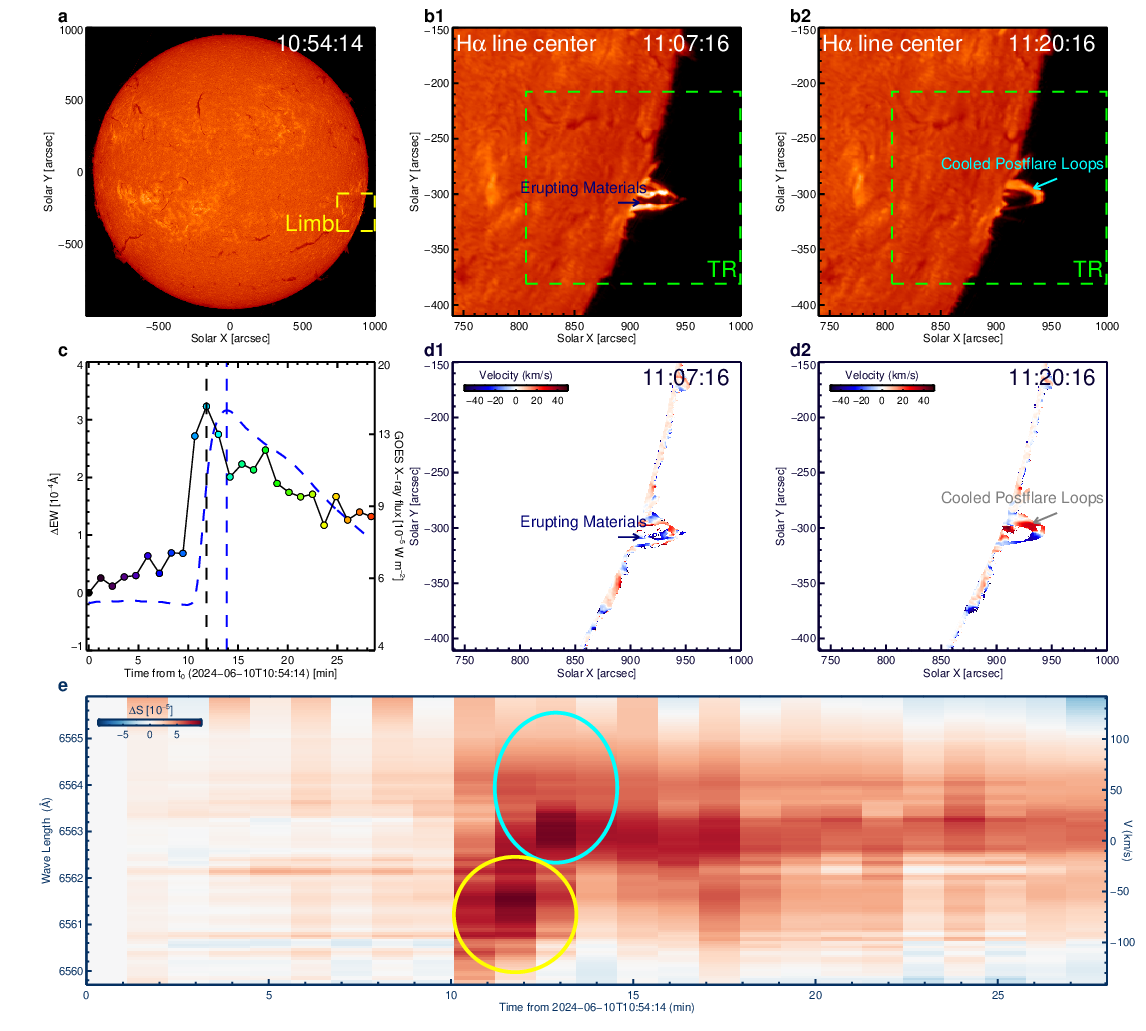}
\caption{CHASE imaging and spectral observations of the event occurred on 2024 Jun 10 (event 3, far-side limb active region eruption). (a)-(e) The same as Figure \ref{fig1}(a1)-(e1) but for event 3. (b1) \& (b2) and (d1) \& (d2) indicate different observation time. The navy and gray arrows mark the erupting materials and postflare loops. An animation is available online, which displays the evolution of event 3. The animation’s duration is 1.3 s.}
\label{fig2}
\end{figure}

The primary distinction between event 1 and event 2 lies in their location: event 1 is an on-disk eruption, while event 2 is a limb eruption. Both events exhibit signatures from flare ribbons, which result in common features in the integrated H$\alpha$ spectrum, namely enhanced emission at the line center and red asymmetry. The key difference stems from the contrasting behavior of the erupting material relative to the solar background. In event 1, the filament acts as an absorbing structure against the bright solar disk, manifesting as an absorption signal in the blue wing. In contrast, the prominence in event 2 appears as an emitting source against the dark sky background, producing an emission asymmetry in the blue wing.

It is noteworthy that, here we only report limb eruptions showing emission in the blue wing. However, limb eruptions exhibiting emission in red wing---indicating an initial line-of-sight velocity component away from the observer---should also exist. Furthermore, such signals caused by the erupting filament are expected to be located at a larger wavelength offset from the line center compared to the red asymmetry typically associated with chromospheric condensation. It should also be noted that on-disk active region filament eruptions are not always associated with a distinct blueshifted absorption signature due to the obscuration by simultaneous dominant flare-ribbon emission (see event 4 and Figure \ref{fig3}(e1) for details). Therefore, distinguishing between on-disk and limb eruption events should rely on the presence or absence of blueshifted/redshifted emission signals---a characteristic feature of limb eruptions.

The diagnostic markers identified here provide a direct interpretative framework for stellar H$\alpha$ spectra. For instance, blueshifted absorption wing in stellar H$\alpha$ profiles during flare events have recently been reported and interpreted as signatures of on-disk eruptive filament materials \citep[e.g.,][]{2022NatAs...6..241N}, while blueshifted emission signatures, characteristic of limb filament eruptions, have also been reported on stars \citep{2024ApJ...961...23N,2025ApJ...978L..32L,2025ApJ...979...93K}. This distinction is crucial for assessing the geometry and kinematics of stellar eruptive events.

\subsection{Front-side vs. Far-side Limb Active Region Eruptions}

Figure \ref{fig2} and the associated animation present a far-side limb active region filament eruption (event 3). Event 3 occurred on June 10, 2024, accompanied by a X1.6 flare. According to GOES, the flare started at 11:03 UT, ended at 11:25 UT, and peaked at 11:08 UT. In the spatially integrated spectral characteristics of the source region, an emission signal first appears in the blue wing (see the yellow ellipse in Figure \ref{fig2}(e)), which rapidly drifts towards the red wing. This indicates that the prominence material in this eruptive event rapidly fell back after its initial eruption toward Earth. Subsequently, an emission enhancement in the H$\alpha$ line center and a red asymmetry are observed (see the cyan ellipse in Figure \ref{fig2}(e)). This is attributed to the expansion and elevation of flare loops beyond the limb during this event, rather than flare ribbons on the far side (see Figure \ref{fig2}(b1) \& (d1) and cooled postflare loops in (b2) \& (d2)), with these loops contributing to this portion of the signal \citep{2024ApJ...974L..13O}.

Both of events 2 and 3 are limb eruptions and exhibit a common limb eruption feature in the integrated H$\alpha$ spectrum: an emission signal in the blue wing. However, event 2 originates from the front side of the solar limb, whereas event 3 originates from the far side. As a result, they show a key observational difference: the front-side limb eruption allows for the direct observation of flare ribbons from the associated active region, while the far-side limb eruption does not. Consequently, in the integrated H$\alpha$ spectrum, a front-side limb eruption first shows the emission in the line center and red asymmetry from the flare ribbons, followed by the blueshifted emission from the off-limb erupting filament. In contrast, for a far-side limb eruption, we first observe the blueshifted prominence emission. The emission in the line center from the flare loops may appear subsequently, or may not be observed at all \citep[e.g., event (8) and (9) in][]{2022ApJ...939...98O}. It is also worth noting that non-eruptive dynamic structures like retracting post-flare loops in limb eruptions produce line center emission with with blueshifted/redshifted absorption \citep{2024ApJ...974L..13O}. This can be easily distinguished from the blueshifted/redshifted emission of limb eruptive filaments. Therefore, the presence or absence of the emission signal in the line center before the appearance of the blueshifted/redshifted prominence emission signal may provide a useful hint for distinguishing between front-side and far-side limb eruption events.

On other stars, the distinction between front-side and far-side eruptions is entirely obscured. Our results suggest a potential diagnostic: the temporal sequence of spectral features in high-cadence stellar H$\alpha$ monitoring. For example, we tentatively suggest that Figure 6(b) in \citet{2024ApJ...961...23N} may show a far-side limb eruption, and Figure 4 in \citet{2024PASJ...76..175I} may show a front-side limb eruption. While challenging due to signal-to-noise and cadence requirements, this diagnostic, if confirmed by larger samples, could offer a method to probe the longitudinal distribution of stellar erupting activity.

\subsection{Active Region vs. Quiet-Sun Region On-disk Eruptions}

\begin{figure}[h]
\centering
\includegraphics [width=0.99\textwidth]{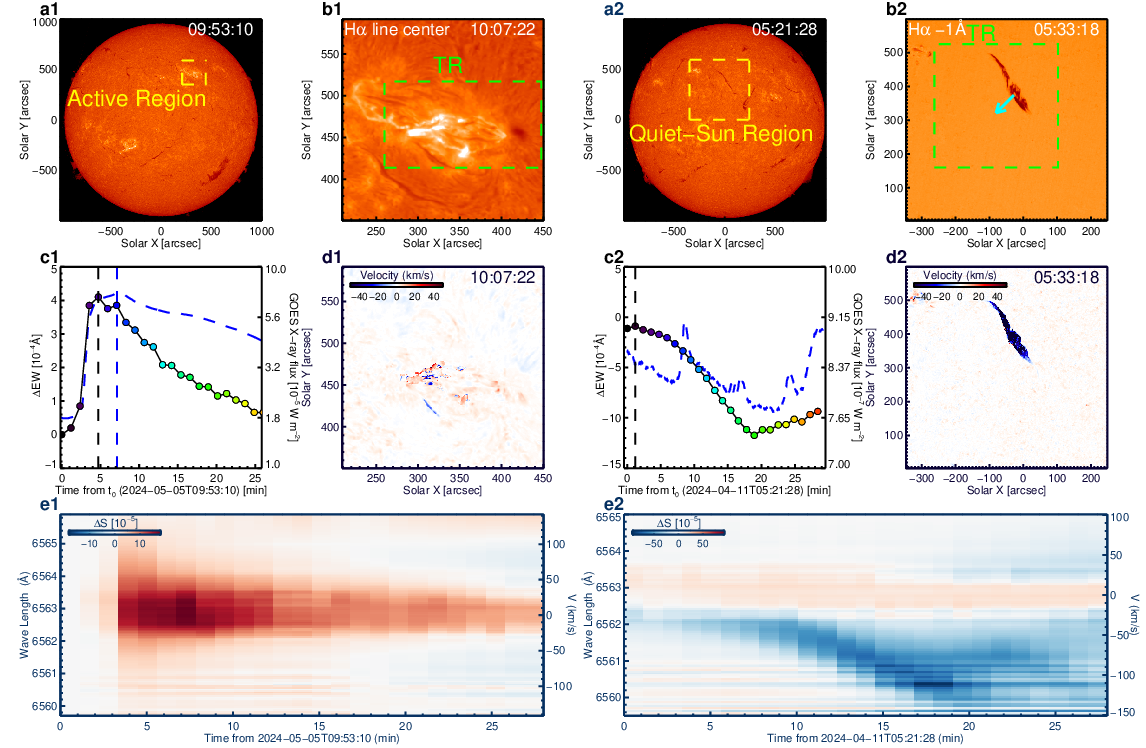}
\caption{CHASE imaging and spectral observations of the events occurred on 2024 May 5 (event 4, on-disk active region eruption) and 2024 Apr 11 (event 5, on-disk quiet-Sun region eruption). (a1)-(e1) The same as Figure \ref{fig1}(a1)-(e1) but for event 4. (a2)-(e2) The same as Figure \ref{fig1}(a1)-(e1) but for event 5. The cyan arrow indicates the erupting direction. An animation is available online, which displays the evolution of events 4 and 5. The animation’s duration is 2.5 s.}
\label{fig3}
\end{figure}

Figure \ref{fig3} and the associated animation present an on-disk active region filament eruption (event 4) and an on-disk quiet-Sun region filament eruption (event 5). Event 4 occurred on May 5, 2024, accompanied by an M7.5 flare. According to GOES records, the flare started at 09:53 UT, ended at 10:19 UT, and peaked at 10:00 UT. Similar to event 1, the spatially integrated spectral characteristics of the source region are primarily dominated by the radiative signatures of flare ribbons (see Figure \ref{fig3}(e1)). These include enhanced emission in the H$\alpha$ line center, an increase in line width, and a red asymmetric emission signal, resulting in an overall approximately symmetric profile. During the subsequent decay phase, the emission intensity gradually decreases while the line profile remains symmetric. Due to the observational time constraints of CHASE, the chosen pre-event reference time $t_0$ does not lie within the pre-flare phase but is already at the beginning of the rise phase. Consequently, toward the end of the decay phase, the intensity falls below the $t_0$ level, manifesting as absorption signals in both wings of the line profile. Furthermore, no distinct blueshifted absorption signature from the erupting filament is observed in this event. In \citet{2025ApJ...993..126L} on major flare events, it is noted that the signals from flare ribbons dominate and mask those from the filament eruption. The $\Delta$EW shows a significant enhancement, peaking before the GOES SXR peak (see Figure \ref{fig3}(c1)).

Event 5 occurred on April 11, 2024. This event is a quiet-Sun region filament eruption, unaccompanied by significant flare ribbon signatures. The spatially integrated spectral characteristics of the source region are primarily characterized by a blueshifted absorption signature from the erupting filament (see Figure \ref{fig3}(e2)). The weak emission signal near the line center results from the disappearance of the filament and a minor flare brightening \citep{2022ApJ...939...98O, 2025ApJ...993..126L}. The SXR flux shows an enhancement, while the $\Delta$EW exhibits a declining trend (see Figure \ref{fig3}(c2)).

The primary distinction between event 4 and event 5 lies in their magnetic environment: event 4 is an active region filament eruption occurring within a region with strong magnetic fields, which is always associated with remarkable flare ribbons, while event 5 originates from a quiet-Sun region characterized by much weaker magnetic fields and flare ribbons (if not absent). Although both events exhibit an emission signal at the H$\alpha$ line center in their integrated spectra, the intensity, origin, and evolution of this signal are fundamentally different. In event 4, the signal is a strong line center emission primarily from flare ribbons and gradually weakens, whereas in event 5, it stems mainly from the disappearance of the filament and is weaker, but progressively increases. Consequently, the intensity and evolution of line center emission signal can serve as an indicator for inferring whether an on-disk filament eruption originates from a strong-field active region or a weak-field quiet-Sun region, which could also be a potential probe to the magnetic environment of stellar filament eruption sources.

Furthermore, a quiet-Sun region filament eruption (event 5) typically displays a blueshifted absorption signature, whereas in active-region events this signature is not always discernible, especially when intense flare ribbons dominate the integrated TR spectrum (event 4). However, event 1---the same type as event 4 but with a stronger M8.7 flare---exhibits a clear absorption signature thanks to the erupting filament’s large projected area; together with the flare ribbon emission, this shapes the integrated spectral characteristics of event 1. This apparent discrepancy between events 1 and 4 arises because the detectability of filament absorption is not determined by flare energy class alone, but by the relative dominance of flare ribbon emission versus erupting filament absorption in the source region, along with geometric factors such as the projected area of the filament, line‑of‑sight velocity, column density, and optical depth.

\begin{figure}[h]
\centering
\includegraphics [width=0.99\textwidth]{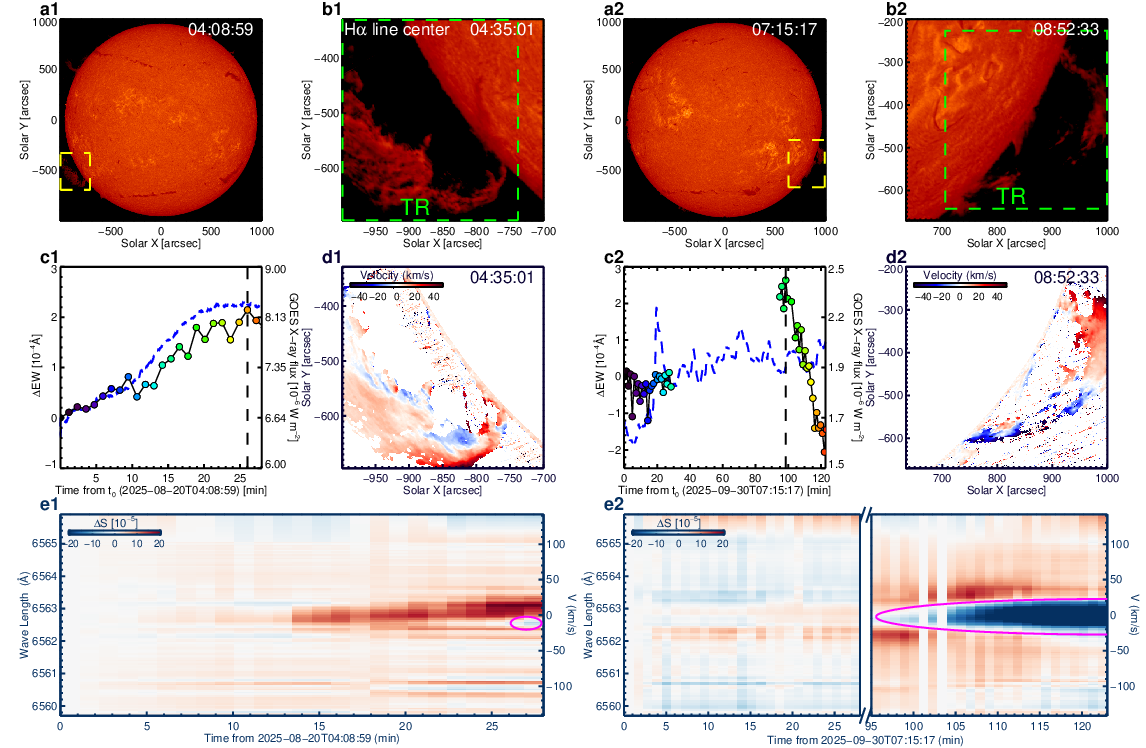}
\caption{CHASE imaging and spectral observations of the events occurred on 2025 Aug 20 (event 6, limb quiet-Sun region eruption) and 2025 Sep 30 (event 7, limb quiet-Sun region eruption). (a1)-(e1) The same as Figure \ref{fig1}(a1)-(e1) but for event 6. (a2)-(e2) The same as Figure \ref{fig1}(a1)-(e1) but for event 7. An animation is available online, which displays the evolution of events 6 and 7. The animation’s duration is 3.7 s.}
\label{fig4}
\end{figure}

\subsection{Active Region vs. Quiet-Sun Region Limb Eruptions}

Figure \ref{fig4} and the associated animation present two limb quiet-Sun region filament (quiescent prominence) eruptions (events 6 and 7). Event 6 occurred on August 20, 2025. This event is a quiescent prominence eruption, unaccompanied by significant flare ribbon signatures. Limited by the observational time constraints of CHASE, our data only capture primarily the rising and acceleration phase of this prominence, with only a little of its subsequent disappearance phase recorded. The spatially integrated spectral characteristics of the source region first manifest as emission at the line center, which subsequently extends slowly toward both wings (see Figure \ref{fig4}(e1)). This is because the quiescent prominence brightens upon activation, enhancing the emission in the line center. Simultaneously, the large-scale prominence expands in the forward and backward directions, leading to the extended emission signals in the wings. Furthermore, it is observed that the signal in the red wing is significantly stronger than that in the blue wing. This indicates that the primary direction of the prominence eruption is away from Earth---a conclusion supported by the calculated velocity field (see Figure \ref{fig4}(d1)). Toward the end of the event, a weak absorption signature appears near the line center (see magenta ellipse in Figure \ref{fig4}(e1)), resulting from the onset of the large-scale prominence's disappearance. The $\Delta$EW shows an increasing trend due to the brightening of the activated prominence (see Figure \ref{fig4}(c1)).

Event 7 occurred on September 30, 2025. This quiescent prominence eruption is also typically unaccompanied by obvious flare ribbon signatures. Limited by the observational time constraints of CHASE, our two sets of data only capture primarily the pre-eruption phase and the final disappearance phase of this prominence, with almost none of the intermediate activation and rise phase recorded. The spatially integrated spectral characteristics initially show a gradual enhancement of the line center emission before the eruption, culminating in a strong absorption signature at the line center during the final phase (see magenta ellipse in Figure \ref{fig4}(e2)). Simultaneously, an emission signal extending slowly toward both wings is observed. Similar to event 6, the initial enhancement in the line center emission is attributed to the activation and subsequent brightening of the quiescent prominence. The extended wing emission results from the expansion of the large-scale prominence structure. The blueshifted emission is initially stronger than the redshifted emission but later becomes weaker. This indicates that the eruptive direction of this prominence is nearly perpendicular to the line of sight, with expanding velocity components both away from and toward the observer---supported by the calculated velocity field, which shows approximately half of the prominence moving away from Earth and the other half moving toward it (see Figure \ref{fig4}(d2)). The intense line center absorption observed at the end of the event is due to the disappearance of the large-scale prominence, leaving behind only a sky background intensity that is weaker than the reference level at $t_0$. The SXR flux shows an overall increase, while the $\Delta$EW first rises and then drops sharply (see Figure \ref{fig4}(c2)). The rise is due to the brightening of the activated prominence, and the subsequent drop results from its disappearance.

Due to the fact that large-scale quiescent prominences typically span both the front and far sides of the solar disk while being suspended off the limb, we therefore do not distinguish between front-side and far-side eruptions in our study. Based on the analysis of event 6 and event 7, we can infer the full sequence of spectral characteristics for a quiescent prominence eruption in H$\alpha$. The characteristic evolution consists of an initial enhancement in the line center emission, which gradually extends toward both wings, followed by a subsequent phase of intense absorption at the line center.

The primary distinction between events 2 \& 3 and events 6 \& 7 is their magnetic environment: events 2 and 3 are active region eruptions occurring within regions of strong magnetic fields, whereas events 6 and 7 originate from quiet-Sun regions with weaker magnetic fields. This difference manifests distinctly in the H$\alpha$ integrated spectra. Limb eruptions from active regions typically exhibit a blueshifted/redshifted emission signal. In contrast, quiet-Sun region eruptions show emission developing in both wings. This divergence arises from their different eruptive dynamics. Active region prominences often erupt as a more collimated, jet-like ejection. The observed Doppler shift in a single wing directly reflects the direction of its initial velocity. Conversely, large-scale quiescent prominences first undergo a prolonged expansion phase, which produces emission in both wings, before a subsequent dominance in one wing may emerge due to the overall eruptive direction. Furthermore, the temporal evolution provides a key discriminant. Quiet-Sun region eruptions generally have a significantly longer duration than their active region counterparts, resulting in a much slower drift of the emission signals in wavelength space.

Another critical distinguishing feature is the spectral profile during the late phase. Quiet-Sun region limb eruptions often culminate in a strong absorption signature at the line center, while active region limb eruptions typically show only an initial line center enhancement (or no line center signal at all, depending on the visibility of flare-ribbon-related structures emissions) and lack such a deep absorption feature. Therefore, the presence of a final-phase line center absorption signal, coupled with initial emission that drifts simultaneously into both wings, could serve as a key diagnostic criterion for distinguishing between limb eruptions originating from quiet-Sun regions and those from active regions.

The prolonged, complex spectral evolution of large-scale quiet-Sun region eruptions presents a unique, multi-feature fingerprint for stellar observations, for which relevant stellar studies are still lacking. Our solar templates suggest that complete, high-cadence monitoring of similar stellar events might reveal this full sequence. Distinguishing these large-scale quiet-Sun region filament eruption events from more frequent, compact active region filament eruptions is vital for accurately quantifying stellar mass-loss rates via eruptions. For applications to stellar observations, two additional considerations are worth noting. First, under the condition of full-disk integration, the contribution from disappearance of such quiescent prominences is likely very small and may even be buried in the noise, let alone a strong signal. However, we do not exclude the possibility that on some stars, the spatial scale and radiative intensity of such prominences could be large enough to produce detectable signal comparable to the entire stellar disk. In such cases, a line center absorption signal might still be observable due to the disappearance of the extreme prominences. Second, if observed over a long duration, an active region limb eruption may also show late-phase line center absorption (e.g., when the region rotates to the far-side or when its post-flare emission declines below pre-eruption levels). However, the central absorption signal arising from the overall evolution of the active region typically appears on a timescale much longer than that caused by an erupting quiescent prominence near the limb. As a result, for stellar observations, it is more appropriate to combine the emission signals in both wings and the evolutionary timescale for a comprehensive diagnosis between the quiet-Sun region limb eruptions and active region eruptions.

\section{Conclusions}\label{4}

In this study, we have analyzed a diverse set of 7 solar filament/prominence eruptions using H$\alpha$ spectroscopic observations from CHASE. By employing the Sun-as-a-star analysis, we integrated spectral profiles from spatially resolved source regions of solar filament eruptions to synthesize the observational signatures that would be recorded for a spatially unresolved stellar filament eruptions. Our primary goal was to investigate whether these integrated H$\alpha$ spectral time series contain diagnostic information that can reveal the source region location and magnetic environment of these eruptions.

Our comparative analysis reveals distinct spectral fingerprints for eruptions originating from different source regions:
\begin{enumerate}
    \item \textbf{On-disk vs. Limb Eruptions:} The presence of a blueshifted/redshifted emission signal is a characteristic feature of off-limb erupting filaments observed. In contrast, on-disk filaments can manifest as blueshifted absorption features. The absence of wing emission signals, combined with line center emission, strongly suggests an on-disk eruption.
    \item \textbf{Front-side vs. Far-side Limb Eruptions:} Front-side limb eruptions, where the associated flare ribbons are visible, first show enhanced line center emission and wing emission asymmetry near the line center, followed by blueshifted/redshifted emission signal from off-limb erupting filaments. Far-side limb eruptions, where the flare site is hidden, exhibit the blueshifted/redshifted emission signal first; line center emission from off-limb flare loops may appear later or be entirely absent.
    \item \textbf{Active Region vs. Quiet-Sun region Eruptions:} The magnetic environment imprints clear signatures on the integrated spectrum. Active region eruptions are characterized by intense, flare-ribbon-induced line center emission with occasional blueshifted absorption for on-disk filaments, and single-wing (blue or red) emission caused by collimated fast ejections for limb filaments. Quiet-Sun region eruptions exhibit much weaker line center brightening, frequently display initial blueshifted absorption signatures for on-disk events, and a prolonged expansion phase resulting in simultaneous emission broadening into both wings for limb events. It should be noted that both active region limb events and quiet‑Sun region limb events may show late‑phase line center absorption, but with a longer timescale for the former; this, together with dual‑wing emission for the latter, could distinguish them in unresolved stellar spectra.
\end{enumerate}

These diagnostic criteria---centered on the presence, temporal evolution, and symmetry of wing emission/absorption signals, the strength and timing of line center emission, and the event timescale---form a foundational framework for interpreting unresolved stellar H$\alpha$ eruption observations. This study demonstrates that, even without spatial resolution, the detailed spectral evolution in H$\alpha$ contains valuable information about the source region of an eruption. This capability is crucial for stellar astrophysics, where it can constrain the location of stellar filament/prominence eruptions (on-disk vs. limb, active vs. quiet-Sun regions), thereby informing models of stellar magnetic activity cycles and coronal mass ejection rates. Our future work will focus on quantifying these diagnostic spectral features to establish robust numerical classification thresholds. It is worth noting that here we only select and present the most representative case from each event category, while the actual situation is often more complex. For example, although events 1 and 4 belong to the same category, they are not entirely identical---event 4 exhibits only partial characteristics typical of this type. As a result we will also apply this diagnostic framework to a larger statistical sample of events and develop automated classification algorithms to maximize the potential of upcoming stellar spectroscopic surveys. Moreover, expanding the event sample to include more cases with diverse TRs and conducting full-disk integrated Sun-as-a-star analyses will be essential for further validating and extending our results to stellar applications. These are important directions that we intend to pursue in subsequent studies.

\begin{acknowledgments}

The authors appreciate the anonymous referee for the constructive comments and valuable suggestions. The data are used courtesy of CHASE and GOES science teams. The CHASE mission is supported by the China National Space Administration (CNSA). The authors are supported by the Strategic Priority Research Program of the Chinese Academy of Sciences (XDB0560000), the National Key R\&D Program of China (2022YFF0503800), the National Natural Science Foundation of China (12273060, 12333009, and 12533010), the Fundamental Research Funds for the Central Universities (KG202506), the Youth Innovation Promotion Association CAS (2023063), postgraduate Research \& Practice Innovation Program of Jiangsu Province (KYCX24\_0183), China’s Space Origins Exploration Program (GJ11020405), and the Specialized Research Fund for State Key Laboratory of Solar Activity and Space Weather.

\end{acknowledgments}

\bibliography{References}{}

@BOOK{2014masu.book.....P,
       author = {{Priest}, Eric},
        title = "{Magnetohydrodynamics of the Sun}",
         year = 2014,
          doi = {10.1017/CBO9781139020732},
       adsurl = {https://ui.adsabs.harvard.edu/abs/2014masu.book.....P}
}

@ARTICLE{2011LRSP....8....1C,
       author = {{Chen}, P.~F.},
        title = "{Coronal Mass Ejections: Models and Their Observational Basis}",
      journal = {Living Reviews in Solar Physics},
         year = 2011,
        month = dec,
       volume = {8},
       number = {1},
          eid = {1},
        pages = {1},
          doi = {10.12942/lrsp-2011-1},
       adsurl = {https://ui.adsabs.harvard.edu/abs/2011LRSP....8....1C}
}

@ARTICLE{2022Atmos..13.1781G,
       author = {{Gopalswamy}, Nat},
        title = "{The Sun and Space Weather}",
      journal = {Atmosphere},
         year = 2022,
        month = oct,
       volume = {13},
       number = {11},
          eid = {1781},
        pages = {1781},
          doi = {10.3390/atmos13111781},
archivePrefix = {arXiv},
       eprint = {2211.06775},
 primaryClass = {astro-ph.SR},
       adsurl = {https://ui.adsabs.harvard.edu/abs/2022Atmos..13.1781G}
}

@PROCEEDINGS{2015ASSL..415.....V,
        title = "{Solar Prominences}",
    booktitle = {Solar Prominences},
         year = 2015,
       editor = {{Vial}, Jean-Claude and {Engvold}, Oddbj{\o}rn},
       series = {Astrophysics and Space Science Library},
       volume = {415},
        month = jan,
          doi = {10.1007/978-3-319-10416-4},
       adsurl = {https://ui.adsabs.harvard.edu/abs/2015ASSL..415.....V}
}

@ARTICLE{2025ApJ...984....4W,
       author = {{Wu}, Zongyin and {Xue}, Zhike and {Yan}, Xiaoli and {Wang}, Jincheng and {Yang}, Liheng and {Xu}, Zhe and {Li}, Qiaoling and {Peng}, Yang and {Yang}, Liping and {Zhou}, Yian and {Zhang}, Xinsheng and {Gong}, Liufan and {Dong}, Qifan and {Wu}, Guotang},
        title = "{The Observations of Magnetic Reconnection during the Interaction Process of Two Active Region Filaments}",
      journal = {\apj},
         year = 2025,
        month = may,
       volume = {984},
       number = {1},
          eid = {4},
        pages = {4},
          doi = {10.3847/1538-4357/adc377},
archivePrefix = {arXiv},
       eprint = {2506.05659},
 primaryClass = {astro-ph.SR},
       adsurl = {https://ui.adsabs.harvard.edu/abs/2025ApJ...984....4W}
}

@ARTICLE{2021ApJ...914...39L,
       author = {{Lynch}, Benjamin J. and {Palmerio}, Erika and {DeVore}, C. Richard and {Kazachenko}, Maria D. and {Dahlin}, Joel T. and {Pomoell}, Jens and {Kilpua}, Emilia K.~J.},
        title = "{Modeling a Coronal Mass Ejection from an Extended Filament Channel. I. Eruption and Early Evolution}",
      journal = {\apj},
         year = 2021,
        month = jun,
       volume = {914},
       number = {1},
          eid = {39},
        pages = {39},
          doi = {10.3847/1538-4357/abf9a9},
archivePrefix = {arXiv},
       eprint = {2104.08643},
 primaryClass = {astro-ph.SR},
       adsurl = {https://ui.adsabs.harvard.edu/abs/2021ApJ...914...39L}
}

@ARTICLE{2022NatAs...6..241N,
       author = {{Namekata}, Kosuke and {Maehara}, Hiroyuki and {Honda}, Satoshi and {Notsu}, Yuta and {Okamoto}, Soshi and {Takahashi}, Jun and {Takayama}, Masaki and {Ohshima}, Tomohito and {Saito}, Tomoki and {Katoh}, Noriyuki and {Tozuka}, Miyako and {Murata}, Katsuhiro L. and {Ogawa}, Futa and {Niwano}, Masafumi and {Adachi}, Ryo and {Oeda}, Motoki and {Shiraishi}, Kazuki and {Isogai}, Keisuke and {Seki}, Daikichi and {Ishii}, Takako T. and {Ichimoto}, Kiyoshi and {Nogami}, Daisaku and {Shibata}, Kazunari},
        title = "{Probable detection of an eruptive filament from a superflare on a solar-type star}",
      journal = {Nature Astronomy},
         year = 2021,
        month = dec,
       volume = {6},
        pages = {241-248},
          doi = {10.1038/s41550-021-01532-8},
archivePrefix = {arXiv},
       eprint = {2112.04808},
 primaryClass = {astro-ph.SR},
       adsurl = {https://ui.adsabs.harvard.edu/abs/2022NatAs...6..241N}
}

@ARTICLE{2024ApJ...961...23N,
       author = {{Namekata}, Kosuke and {Airapetian}, Vladimir S. and {Petit}, Pascal and {Maehara}, Hiroyuki and {Ikuta}, Kai and {Inoue}, Shun and {Notsu}, Yuta and {Paudel}, Rishi R. and {Arzoumanian}, Zaven and {Avramova-Boncheva}, Antoaneta A. and {Gendreau}, Keith and {Jeffers}, Sandra V. and {Marsden}, Stephen and {Morin}, Julien and {Neiner}, Coralie and {Vidotto}, Aline A. and {Shibata}, Kazunari},
        title = "{Multiwavelength Campaign Observations of a Young Solar-type Star, EK Draconis. I. Discovery of Prominence Eruptions Associated with Superflares}",
      journal = {\apj},
         year = 2024,
        month = jan,
       volume = {961},
       number = {1},
          eid = {23},
        pages = {23},
          doi = {10.3847/1538-4357/ad0b7c},
archivePrefix = {arXiv},
       eprint = {2311.07380},
 primaryClass = {astro-ph.SR},
       adsurl = {https://ui.adsabs.harvard.edu/abs/2024ApJ...961...23N}
}

@ARTICLE{2025ApJ...978L..32L,
       author = {{Lu}, Hong-Peng and {Tian}, Hui and {Zhang}, Li-Yun and {Chen}, He-Chao and {Li}, Ying and {Yang}, Zi-Hao and {Wang}, Jia-Sheng and {Zhang}, Jia-Le and {Sun}, Zheng},
        title = "{An Extreme Stellar Prominence Eruption Observed by LAMOST Time-domain Spectroscopy}",
      journal = {\apjl},
         year = 2025,
        month = jan,
       volume = {978},
       number = {2},
          eid = {L32},
        pages = {L32},
          doi = {10.3847/2041-8213/ad93cc},
archivePrefix = {arXiv},
       eprint = {2411.11076},
 primaryClass = {astro-ph.SR},
       adsurl = {https://ui.adsabs.harvard.edu/abs/2025ApJ...978L..32L}
}

@ARTICLE{2016SoPh..291.1761H,
       author = {{Harra}, Louise K. and {Schrijver}, Carolus J. and {Janvier}, Miho and {Toriumi}, Shin and {Hudson}, Hugh and {Matthews}, Sarah and {Woods}, Magnus M. and {Hara}, Hirohisa and {Guedel}, Manuel and {Kowalski}, Adam and {Osten}, Rachel and {Kusano}, Kanya and {Lueftinger}, Theresa},
        title = "{The Characteristics of Solar X-Class Flares and CMEs: A Paradigm for Stellar Superflares and Eruptions?}",
      journal = {\solphys},
         year = 2016,
        month = aug,
       volume = {291},
       number = {6},
        pages = {1761-1782},
          doi = {10.1007/s11207-016-0923-0},
       adsurl = {https://ui.adsabs.harvard.edu/abs/2016SoPh..291.1761H}
}

@ARTICLE{2022ApJ...933..209N,
       author = {{Namekata}, Kosuke and {Ichimoto}, Kiyoshi and {Ishii}, Takako T. and {Shibata}, Kazunari},
        title = "{Sun-as-a-star Analysis of H{\ensuremath{\alpha}} Spectra of a Solar Flare Observed by SMART/SDDI: Time Evolution of Red Asymmetry and Line Broadening}",
      journal = {\apj},
         year = 2022,
        month = jul,
       volume = {933},
       number = {2},
          eid = {209},
        pages = {209},
          doi = {10.3847/1538-4357/ac75cd},
archivePrefix = {arXiv},
       eprint = {2206.01395},
 primaryClass = {astro-ph.SR},
       adsurl = {https://ui.adsabs.harvard.edu/abs/2022ApJ...933..209N}
}

@ARTICLE{2022ApJ...939...98O,
       author = {{Otsu}, Takato and {Asai}, Ayumi and {Ichimoto}, Kiyoshi and {Ishii}, Takako T. and {Namekata}, Kosuke},
        title = "{Sun-as-a-star Analyses of Various Solar Active Events Using H{\ensuremath{\alpha}} Spectral Images Taken by SMART/SDDI}",
      journal = {\apj},
         year = 2022,
        month = nov,
       volume = {939},
       number = {2},
          eid = {98},
        pages = {98},
          doi = {10.3847/1538-4357/ac9730},
archivePrefix = {arXiv},
       eprint = {2210.02819},
 primaryClass = {astro-ph.SR},
       adsurl = {https://ui.adsabs.harvard.edu/abs/2022ApJ...939...98O}
}

@ARTICLE{2025ApJ...993..126L,
       author = {{Liu}, Xiaofeng and {Hou}, Yijun and {Li}, Ying and {Qiu}, Ye and {Li}, Ting and {Cai}, Yingjie and {Rao}, Shihao and {Zhang}, Junyi and {Li}, Chuan},
        title = "{Sun-as-a-star Analysis of the Solar Eruption Source Region Using H{\ensuremath{\alpha}} Spectroscopic Observations of CHASE}",
      journal = {\apj},
         year = 2025,
        month = nov,
       volume = {993},
       number = {1},
          eid = {126},
        pages = {126},
          doi = {10.3847/1538-4357/ae0743},
archivePrefix = {arXiv},
       eprint = {2508.17762},
 primaryClass = {astro-ph.SR},
       adsurl = {https://ui.adsabs.harvard.edu/abs/2025ApJ...993..126L}
}

@ARTICLE{2024ApJ...964...75O,
       author = {{Otsu}, Takato and {Asai}, Ayumi},
        title = "{Multiwavelength Sun-as-a-star Analysis of the M8.7 Flare on 2022 October 2 Using H{\ensuremath{\alpha}} and EUV Spectra Taken by SMART/SDDI and SDO/EVE}",
      journal = {\apj},
         year = 2024,
        month = mar,
       volume = {964},
       number = {1},
          eid = {75},
        pages = {75},
          doi = {10.3847/1538-4357/ad24ec},
archivePrefix = {arXiv},
       eprint = {2402.00589},
 primaryClass = {astro-ph.SR},
       adsurl = {https://ui.adsabs.harvard.edu/abs/2024ApJ...964...75O}
}

@ARTICLE{2024ApJ...966...45M,
       author = {{Ma}, Y.~L. and {Lao}, Q.~H. and {Cheng}, X. and {Wang}, B.~T. and {Zhao}, Z.~H. and {Rao}, S.~H. and {Li}, C. and {Ding}, M.~D.},
        title = "{Sun-as-a-star Study of an X-class Solar Flare with Spectroscopic Observations of CHASE}",
      journal = {\apj},
         year = 2024,
        month = may,
       volume = {966},
       number = {1},
          eid = {45},
        pages = {45},
          doi = {10.3847/1538-4357/ad3446},
archivePrefix = {arXiv},
       eprint = {2403.09011},
 primaryClass = {astro-ph.SR},
       adsurl = {https://ui.adsabs.harvard.edu/abs/2024ApJ...966...45M}
}

@ARTICLE{2024ApJ...974L..13O,
       author = {{Otsu}, Takato and {Asai}, Ayumi and {Ikuta}, Kai and {Shibata}, Kazunari},
        title = "{Sun-as-a-star Analysis of the X1.6 Flare on 2023 August 5: Dynamics of Postflare Loops in Spatially Integrated Observational Data}",
      journal = {\apjl},
         year = 2024,
        month = oct,
       volume = {974},
       number = {1},
          eid = {L13},
        pages = {L13},
          doi = {10.3847/2041-8213/ad7a70},
archivePrefix = {arXiv},
       eprint = {2409.07630},
 primaryClass = {astro-ph.SR},
       adsurl = {https://ui.adsabs.harvard.edu/abs/2024ApJ...974L..13O}
}

@ARTICLE{2022SCPMA..6589602L,
       author = {{Li}, Chuan and {Fang}, Cheng and {Li}, Zhen and {Ding}, MingDe and {Chen}, PengFei and {Qiu}, Ye and {You}, Wei and {Yuan}, Yuan and {An}, MinJie and {Tao}, HongJiang and {Li}, XianSheng and {Chen}, Zhe and {Liu}, Qiang and {Mei}, Gui and {Yang}, Liang and {Zhang}, Wei and {Cheng}, WeiQiang and {Chen}, JianXin and {Chen}, ChangYa and {Gu}, Qiang and {Huang}, QingLong and {Liu}, MingXing and {Han}, ChengShan and {Xin}, HongWei and {Chen}, ChangZheng and {Ni}, YiWei and {Wang}, WenBo and {Rao}, ShiHao and {Li}, HaiTang and {Lu}, Xi and {Wang}, Wei and {Lin}, Jun and {Jiang}, YiXian and {Meng}, LingJie and {Zhao}, Jian},
        title = "{The Chinese H{\ensuremath{\alpha}} Solar Explorer (CHASE) mission: An overview}",
      journal = {Science China Physics, Mechanics, and Astronomy},
         year = 2022,
        month = aug,
       volume = {65},
       number = {8},
          eid = {289602},
        pages = {289602},
          doi = {10.1007/s11433-022-1893-3},
archivePrefix = {arXiv},
       eprint = {2205.05962},
 primaryClass = {astro-ph.SR},
       adsurl = {https://ui.adsabs.harvard.edu/abs/2022SCPMA..6589602L}
}

@ARTICLE{2022SCPMA..6589603Q,
       author = {{Qiu}, Ye and {Rao}, ShiHao and {Li}, Chuan and {Fang}, Cheng and {Ding}, MingDe and {Li}, Zhen and {Ni}, YiWei and {Wang}, WenBo and {Hong}, Jie and {Hao}, Qi and {Dai}, Yu and {Chen}, PengFei and {Wan}, XiaoSheng and {Xu}, Zhi and {You}, Wei and {Yuan}, Yuan and {Tao}, HongJiang and {Li}, XianSheng and {He}, YuKun and {Liu}, Qiang},
        title = "{Calibration procedures for the CHASE/HIS science data}",
      journal = {Science China Physics, Mechanics, and Astronomy},
         year = 2022,
        month = aug,
       volume = {65},
       number = {8},
          eid = {289603},
        pages = {289603},
          doi = {10.1007/s11433-022-1900-5},
archivePrefix = {arXiv},
       eprint = {2205.06075},
 primaryClass = {astro-ph.SR},
       adsurl = {https://ui.adsabs.harvard.edu/abs/2022SCPMA..6589603Q}
}

@ARTICLE{2024ApJ...961L..30Q,
       author = {{Qiu}, Ye and {Li}, Chuan and {Guo}, Yang and {Li}, Zhen and {Ding}, Mingde and {Kong}, Linggao},
        title = "{Three-dimensional Velocity Fields of the Solar Filament Eruptions Detected by CHASE}",
      journal = {\apjl},
         year = 2024,
        month = feb,
       volume = {961},
       number = {2},
          eid = {L30},
        pages = {L30},
          doi = {10.3847/2041-8213/ad1e4f},
archivePrefix = {arXiv},
       eprint = {2401.16730},
 primaryClass = {astro-ph.SR},
       adsurl = {https://ui.adsabs.harvard.edu/abs/2024ApJ...961L..30Q}
}

@ARTICLE{2024PASJ...76..175I,
       author = {{Inoue}, Shun and {Enoto}, Teruaki and {Namekata}, Kosuke and {Notsu}, Yuta and {Honda}, Satoshi and {Maehara}, Hiroyuki and {Zhang}, Jiale and {Lu}, Hong-Peng and {Uchida}, Hiroyuki and {Tsuru}, Takeshi Go and {Nogami}, Daisaku and {Shibata}, Kazunari},
        title = "{Multiwavelength observation of an active M-dwarf star EV Lacertae and its stellar flare accompanied by a delayed prominence eruption}",
      journal = {\pasj},
         year = 2024,
        month = apr,
       volume = {76},
       number = {2},
        pages = {175-190},
          doi = {10.1093/pasj/psae001},
archivePrefix = {arXiv},
       eprint = {2401.00399},
 primaryClass = {astro-ph.SR},
       adsurl = {https://ui.adsabs.harvard.edu/abs/2024PASJ...76..175I}
}

@ARTICLE{2010SSRv..151..333M,
       author = {{Mackay}, D.~H. and {Karpen}, J.~T. and {Ballester}, J.~L. and {Schmieder}, B. and {Aulanier}, G.},
        title = "{Physics of Solar Prominences: II{\textemdash}Magnetic Structure and Dynamics}",
      journal = {\ssr},
         year = 2010,
        month = apr,
       volume = {151},
       number = {4},
        pages = {333-399},
          doi = {10.1007/s11214-010-9628-0},
archivePrefix = {arXiv},
       eprint = {1001.1635},
 primaryClass = {astro-ph.SR},
       adsurl = {https://ui.adsabs.harvard.edu/abs/2010SSRv..151..333M}
}

@ARTICLE{2020RAA....20..166C,
       author = {{Chen}, Peng-Fei and {Xu}, Ao-Ao and {Ding}, Ming-De},
        title = "{Some interesting topics provoked by the solar filament research in the past decade}",
      journal = {Research in Astronomy and Astrophysics},
         year = 2020,
        month = oct,
       volume = {20},
       number = {10},
          eid = {166},
        pages = {166},
          doi = {10.1088/1674-4527/20/10/166},
archivePrefix = {arXiv},
       eprint = {2010.02462},
 primaryClass = {astro-ph.SR},
       adsurl = {https://ui.adsabs.harvard.edu/abs/2020RAA....20..166C}
}

@ARTICLE{1943MNRAS.103..244N,
       author = {{Newton}, H.~W.},
        title = "{Solar flares and magnetic storms}",
      journal = {\mnras},
         year = 1943,
        month = jan,
       volume = {103},
        pages = {244},
          doi = {10.1093/mnras/103.5.244},
       adsurl = {https://ui.adsabs.harvard.edu/abs/1943MNRAS.103..244N}
}

@ARTICLE{2011JGRA..116.4104W,
       author = {{Wang}, Yuming and {Chen}, Caixia and {Gui}, Bin and {Shen}, Chenglong and {Ye}, Pinzhong and {Wang}, S.},
        title = "{Statistical study of coronal mass ejection source locations: Understanding CMEs viewed in coronagraphs}",
      journal = {Journal of Geophysical Research (Space Physics)},
         year = 2011,
        month = apr,
       volume = {116},
       number = {A4},
          eid = {A04104},
        pages = {A04104},
          doi = {10.1029/2010JA016101},
archivePrefix = {arXiv},
       eprint = {1101.0641},
 primaryClass = {astro-ph.SR},
       adsurl = {https://ui.adsabs.harvard.edu/abs/2011JGRA..116.4104W}
}

@ARTICLE{2015AdSpR..55.2745S,
       author = {{Schrijver}, Carolus J. and {Kauristie}, Kirsti and {Aylward}, Alan D. and {Denardini}, Clezio M. and {Gibson}, Sarah E. and {Glover}, Alexi and {Gopalswamy}, Nat and {Grande}, Manuel and {Hapgood}, Mike and {Heynderickx}, Daniel and {Jakowski}, Norbert and {Kalegaev}, Vladimir V. and {Lapenta}, Giovanni and {Linker}, Jon A. and {Liu}, Siqing and {Mandrini}, Cristina H. and {Mann}, Ian R. and {Nagatsuma}, Tsutomu and {Nandy}, Dibyendu and {Obara}, Takahiro and {Paul O'Brien}, T. and {Onsager}, Terrance and {Opgenoorth}, Hermann J. and {Terkildsen}, Michael and {Valladares}, Cesar E. and {Vilmer}, Nicole},
        title = "{Understanding space weather to shield society: A global road map for 2015-2025 commissioned by COSPAR and ILWS}",
      journal = {Advances in Space Research},
         year = 2015,
        month = jun,
       volume = {55},
       number = {12},
        pages = {2745-2807},
          doi = {10.1016/j.asr.2015.03.023},
archivePrefix = {arXiv},
       eprint = {1503.06135},
 primaryClass = {physics.space-ph},
       adsurl = {https://ui.adsabs.harvard.edu/abs/2015AdSpR..55.2745S}
}

@ARTICLE{2020JGRA..12527530W,
       author = {{Wang}, Jingjing and {Hoeksema}, J. Todd and {Liu}, Siqing},
        title = "{The Deflection of Coronal Mass Ejections by the Ambient Coronal Magnetic Field Configuration}",
      journal = {Journal of Geophysical Research (Space Physics)},
         year = 2020,
        month = aug,
       volume = {125},
       number = {8},
          eid = {e27530},
        pages = {e27530},
          doi = {10.1029/2019JA027530},
archivePrefix = {arXiv},
       eprint = {1909.06410},
 primaryClass = {astro-ph.SR},
       adsurl = {https://ui.adsabs.harvard.edu/abs/2020JGRA..12527530W}
}

@ARTICLE{2023ApJ...953...68L,
       author = {{Lu}, Hong-peng and {Tian}, Hui and {Chen}, He-chao and {Xu}, Yu and {Hou}, Zhen-yong and {Bai}, Xian-yong and {Tan}, Guang-yu and {Yang}, Zi-hao and {Ren}, Jie},
        title = "{Full Velocities and Propagation Directions of Coronal Mass Ejections Inferred from Simultaneous Full-disk Imaging and Sun-as-a-star Spectroscopic Observations}",
      journal = {\apj},
         year = 2023,
        month = aug,
       volume = {953},
       number = {1},
          eid = {68},
        pages = {68},
          doi = {10.3847/1538-4357/acd6a1},
archivePrefix = {arXiv},
       eprint = {2305.08765},
 primaryClass = {astro-ph.SR},
       adsurl = {https://ui.adsabs.harvard.edu/abs/2023ApJ...953...68L}
}

@ARTICLE{2025ApJ...978...41S,
       author = {{Sahade}, Abril and {Vourlidas}, Angelos and {Mac Cormack}, Cecilia},
        title = "{Analysis of Solar Eruptions Deflecting in the Low Corona: Influence of the Magnetic Environment}",
      journal = {\apj},
         year = 2025,
        month = jan,
       volume = {978},
       number = {1},
          eid = {41},
        pages = {41},
          doi = {10.3847/1538-4357/ad96ba},
archivePrefix = {arXiv},
       eprint = {2411.11599},
 primaryClass = {astro-ph.SR},
       adsurl = {https://ui.adsabs.harvard.edu/abs/2025ApJ...978...41S}
}

@article{2018A&A...619A.100H,
  author = {{Hou}, Y.~J. and {Zhang}, J. and {Li}, T. and {Yang}, S.~H. and {Li}, X.~H.},
        title = "{Eruption of a multi-flux-rope system in solar active region 12673 leading to the two largest flares in Solar Cycle 24}",
      journal = {\aap},
         year = 2018,
        month = nov,
       volume = {619},
          eid = {A100},
        pages = {A100},
          doi = {10.1051/0004-6361/201732530},
archivePrefix = {arXiv},
       eprint = {1808.06795},
 primaryClass = {astro-ph.SR},
       adsurl = {https://ui.adsabs.harvard.edu/abs/2018A&A...619A.100H}
}

@article{Hou2023Apj,
author = {Hou, Yijun and Li, Chuan and Li, Ting and Su, Jiangtao and Qiu, Ye and Yang, Shuhong and Liheng, Yang and Li, Leping and Guo, Yilin and Hou, Zhengyong and Song, Qiao and Bai, Xianyong and Zhou, G. and Ding, Mingde and Gan, Weiqun and Deng, Yuanyong},
year = {2023},
month = {12},
pages = {69},
title = {Partial Eruption of Solar Filaments. I. Configuration and Formation of Double-decker Filaments},
volume = {959},
journal = {The Astrophysical Journal},
doi = {10.3847/1538-4357/ad08bd}
}

@ARTICLE{1995ApJ...443..818L,
       author = {{Low}, B.~C. and {Hundhausen}, J.~R.},
        title = "{Magnetostatic Structures of the Solar Corona. II. The Magnetic Topology of Quiescent Prominences}",
      journal = {\apj},
         year = 1995,
        month = apr,
       volume = {443},
        pages = {818},
          doi = {10.1086/175572},
       adsurl = {https://ui.adsabs.harvard.edu/abs/1995ApJ...443..818L}
}

@ARTICLE{2020A&A...640A.101H,
       author = {{Hou}, Y.~J. and {Li}, T. and {Song}, Z.~P. and {Zhang}, J.},
        title = "{External reconnection and resultant reconfiguration of overlying magnetic fields during sympathetic eruptions of two filaments}",
      journal = {\aap},
         year = 2020,
        month = aug,
       volume = {640},
          eid = {A101},
        pages = {A101},
          doi = {10.1051/0004-6361/202038348},
archivePrefix = {arXiv},
       eprint = {2006.06191},
 primaryClass = {astro-ph.SR},
       adsurl = {https://ui.adsabs.harvard.edu/abs/2020A&A...640A.101H}
}

@ARTICLE{2025ApJ...981..139Y,
       author = {{Yan}, Xiaoli and {Xue}, Zhike and {Wang}, Jincheng and {Chen}, Pengfei and {Ji}, Kaifan and {Xia}, Chun and {Yang}, Liheng and {Kong}, Defang and {Xu}, Zhe and {Zhou}, Yian and {Li}, Qiaoling},
        title = "{Simultaneous Existence of Oscillations, Counterstreaming Flows, and Mass Injections in Solar Quiescent Prominences}",
      journal = {\apj},
         year = 2025,
        month = mar,
       volume = {981},
       number = {2},
          eid = {139},
        pages = {139},
          doi = {10.3847/1538-4357/adb39e},
archivePrefix = {arXiv},
       eprint = {2502.04114},
 primaryClass = {astro-ph.SR},
       adsurl = {https://ui.adsabs.harvard.edu/abs/2025ApJ...981..139Y}
}

@ARTICLE{2025ApJ...979...93K,
       author = {{Kajikiya}, Yuto and {Namekata}, Kosuke and {Notsu}, Yuta and {Maehara}, Hiroyuki and {Sato}, Bunei and {Nogami}, Daisaku},
        title = "{High-time-cadence Spectroscopy and Photometry of Stellar Flares on M dwarf YZ Canis Minoris with the Seimei Telescope and TESS. I. Discovery of Rapid and Short-duration Prominence Eruptions}",
      journal = {\apj},
         year = 2025,
        month = jan,
       volume = {979},
       number = {1},
          eid = {93},
        pages = {93},
          doi = {10.3847/1538-4357/ad91b9},
archivePrefix = {arXiv},
       eprint = {2411.08462},
 primaryClass = {astro-ph.SR},
       adsurl = {https://ui.adsabs.harvard.edu/abs/2025ApJ...979...93K}
}

@ARTICLE{2026NatAs..10...64N,
       author = {{Namekata}, Kosuke and {France}, Kevin and {Chae}, Jongchul and {Airapetian}, Vladimir S. and {Kowalski}, Adam and {Notsu}, Yuta and {Young}, Peter R. and {Honda}, Satoshi and {Kang}, Soosang and {Kang}, Juhyung and {Lee}, Kyeore and {Maehara}, Hiroyuki and {Lee}, Kyoung-Sun and {Tamburri}, Cole and {Ohshima}, Tomohito and {Takayama}, Masaki and {Shibata}, Kazunari},
        title = "{Discovery of multi-temperature coronal mass ejection signatures from a young solar analogue}",
      journal = {Nature Astronomy},
         year = 2026,
        month = jan,
       volume = {10},
        pages = {64-75},
          doi = {10.1038/s41550-025-02691-8},
       adsurl = {https://ui.adsabs.harvard.edu/abs/2026NatAs..10...64N}
}

@ARTICLE{2024LRSP...21....1K,
       author = {{Kowalski}, Adam F.},
        title = "{Stellar flares}",
      journal = {Living Reviews in Solar Physics},
         year = 2024,
        month = dec,
       volume = {21},
       number = {1},
          eid = {1},
        pages = {1},
          doi = {10.1007/s41116-024-00039-4},
archivePrefix = {arXiv},
       eprint = {2402.07885},
 primaryClass = {astro-ph.SR},
       adsurl = {https://ui.adsabs.harvard.edu/abs/2024LRSP...21....1K}
}

@ARTICLE{2024A&A...682A..46P,
       author = {{Pietrow}, A.~G.~M. and {Cretignier}, M. and {Druett}, M.~K. and {Alvarado-G{\'o}mez}, J.~D. and {Hofmeister}, S.~J. and {Verma}, M. and {Kamlah}, R. and {Baratella}, M. and {Amazo-G{\'o}mez}, E.~M. and {Kontogiannis}, I. and {Dineva}, E. and {Warmuth}, A. and {Denker}, C. and {Poppenhaeger}, K. and {Andriienko}, O. and {Dumusque}, X. and {L{\"o}fdahl}, M.~G.},
        title = "{A comparative study of two X2.2 and X9.3 solar flares observed with HARPS-N. Reconciling Sun-as-a-star spectroscopy and high-spatial resolution solar observations in the context of the solar-stellar connection}",
      journal = {\aap},
         year = 2024,
        month = feb,
       volume = {682},
          eid = {A46},
        pages = {A46},
          doi = {10.1051/0004-6361/202347895},
archivePrefix = {arXiv},
       eprint = {2309.03373},
 primaryClass = {astro-ph.SR},
       adsurl = {https://ui.adsabs.harvard.edu/abs/2024A&A...682A..46P}
}

@ARTICLE{2014LRSP...11....1P,
       author = {{Parenti}, Susanna},
        title = "{Solar Prominences: Observations}",
      journal = {Living Reviews in Solar Physics},
         year = 2014,
        month = dec,
       volume = {11},
       number = {1},
          eid = {1},
        pages = {1},
          doi = {10.12942/lrsp-2014-1},
       adsurl = {https://ui.adsabs.harvard.edu/abs/2014LRSP...11....1P}
}

@ARTICLE{2022ApJS..260...36Y,
       author = {{Yang}, Zihao and {Tian}, Hui and {Bai}, Xianyong and {Chen}, Yajie and {Guo}, Yang and {Zhu}, Yingjie and {Cheng}, Xin and {Gao}, Yuhang and {Xu}, Yu and {Chen}, Hechao and {Zhang}, Jiale},
        title = "{Can We Detect Coronal Mass Ejections through Asymmetries of Sun-as-a-star Extreme-ultraviolet Spectral Line Profiles?}",
      journal = {\apjs},
         year = 2022,
        month = jun,
       volume = {260},
       number = {2},
          eid = {36},
        pages = {36},
          doi = {10.3847/1538-4365/ac6607},
archivePrefix = {arXiv},
       eprint = {2204.03683},
 primaryClass = {astro-ph.SR},
       adsurl = {https://ui.adsabs.harvard.edu/abs/2022ApJS..260...36Y}
}

@ARTICLE{2022ApJ...931...76X,
       author = {{Xu}, Yu and {Tian}, Hui and {Hou}, Zhenyong and {Yang}, Zihao and {Gao}, Yuhang and {Bai}, Xianyong},
        title = "{Sun-as-a-star Spectroscopic Observations of the Line-of-sight Velocity of a Solar Eruption on 2021 October 28}",
      journal = {\apj},
         year = 2022,
        month = jun,
       volume = {931},
       number = {2},
          eid = {76},
        pages = {76},
          doi = {10.3847/1538-4357/ac69d5},
archivePrefix = {arXiv},
       eprint = {2204.11722},
 primaryClass = {astro-ph.SR},
       adsurl = {https://ui.adsabs.harvard.edu/abs/2022ApJ...931...76X}
}

@ARTICLE{2026ApJ...997..242C,
       author = {{Cheng}, Zhixun and {Song}, Anchuan and {Zhou}, Guiping and {Wang}, Yuming and {Tian}, Hui and {Li}, Xiaolei and {Liu}, Jifeng and {Wang}, Jingxiu},
        title = "{Detection of Stellar Mass Ejections through Extreme Ultraviolet Spectral Lines}",
      journal = {\apj},
         year = 2026,
        month = feb,
       volume = {997},
       number = {2},
          eid = {242},
        pages = {242},
          doi = {10.3847/1538-4357/ae29a9},
       adsurl = {https://ui.adsabs.harvard.edu/abs/2026ApJ...997..242C}
}
\bibliographystyle{aasjournal}

\end{document}